\documentclass[conference]{IEEEtran}

\usepackage[T1]{fontenc}
\usepackage{amssymb}
\usepackage{amsmath}
\usepackage{algorithm}
\usepackage{algorithmic}

\usepackage{graphicx}
\usepackage{epstopdf}

\usepackage{subfigure}
\usepackage{cite}

\begin{document}
\title{A Hierarchical Rate Splitting Strategy for FDD Massive MIMO under Imperfect CSIT \\}
\author{\IEEEauthorblockN{Mingbo Dai\IEEEauthorrefmark{1}, Bruno Clerckx\IEEEauthorrefmark{1}\IEEEauthorrefmark{2}, David Gesbert\IEEEauthorrefmark{3} and Giuseppe Caire\IEEEauthorrefmark{4}}
\IEEEauthorblockA{\IEEEauthorrefmark{1}Department of Electrical and Electronic Engineering, Imperial College London, UK\\\IEEEauthorrefmark{2}School of Electrical Engineering, Korea University, Seoul, Korea\\\IEEEauthorrefmark{3}Mobile Communications Department, EURECOM, 06560 Sophia Antipolis, France\\\IEEEauthorrefmark{4}Department of Telecommunication Systems, Technical University of Berlin, Berlin, Germany\\Email:\{m.dai13,b.clerckx\}@imperial.ac.uk, david.gesbert@eurecom.fr and caire@tu-berlin.de}}


\maketitle

\begin{abstract}
In a multiuser MIMO broadcast channel, the rate performance is affected by the multiuser interference when the Channel State Information at the Transmitter (CSIT) is imperfect. To tackle the interference problem, a Rate-Splitting (RS) approach has been proposed recently, which splits one user's message into a common and a private part, and superimposes the common message on top of the private messages. The common message is drawn from a public codebook and should be decoded by all users. In this paper, we propose a novel and general framework, denoted as Hierarchical Rate Splitting (HRS), that is particularly suited to FDD massive MIMO systems. HRS simultaneously transmits private messages intended to each user and two kinds of common messages that can be decoded by all users and by a subset of users, respectively. We analyse the asymptotic sum rate of HRS under imperfect CSIT. A closed-form power allocation is derived which provides insights into the effects of system parameters. Finally, simulation results validate the significant sum rate gain of HRS over various baselines.\footnote{This work was partially supported by the Seventh Framework Programme for Research of the European Commission under grant number HARP-318489.} 
\end{abstract}


\IEEEpeerreviewmaketitle

\section{INTRODUCTION}

In MIMO wireless networks, the rate performance is affected by the multiuser interference when the Channel State Information at the Transmitter (CSIT) is imperfect \cite{bruno2013}. To tackle the detrimental effect of multiuser interference, a Rate-Splitting (RS) approach was recently proposed \cite[Lemma 2]{yang2013}. Specifically, we can split one selected user's message (e.g., user $1$) into a common part $(s_c)$ and a private part $(s_1)$, where the common message is drawn from a public codebook and should be decoded by all users with zero error probability. The private message $(s_1)$ and the private messages $(s_{k|k \ne 1})$ intended to other users are transmitted via Zero-Forcing (ZF) beamforming using a fraction of the total power while the common message $(s_c)$ is superimposed on top of the ZF-precoded private messages using the residual power. At the receiver side, the common message is decoded by treating all the private messages as noise. After removing the decoded common message from the received signal by Successive Interference Cancellation (SIC), each user decodes their own private messages.
When the CSIT error variance $\tau^2$ decays with signal-to-noise ratio ($P$) as $O(P^{-\delta})$ for some constant $0 \le \delta < 1$, the sum DoF of RS is $1 + \delta$, which is strictly larger than the $2\delta$ that is achieved with conventional multiuser broadcasting strategies (e.g., ZF).

At finite Signal-to-Noise-Ratio (SNR), by optimizing the transmit beamforming and power allocation parameters for both RS and conventional multiuser broadcasting (BC) scheme, the RS approach shows significant sum rate gain over the conventional BC\cite{hamdi2015fir}. In the context of a two-user MISO broadcast channel with quantized CSIT, \cite{hao2015} has also confirmed the rate benefits provided by the transmission of a common message in a RS strategy over conventional multiuser BC schemes. Both the optimization method proposed in \cite{hamdi2015fir} and the analysis in \cite{hao2015} are hard to extend to multiuser massive MIMO systems, where attaining accurate CSIT gets particularly challenging as the number of transmit antennas increases.

When it comes to designing precoders on the basis of reduced CSI feedback, a two-tier precoder relying on both short- and long-term CSIT has been proposed by several authors \cite{adhikary2013, Jaehyun2014, Kim2014, chen2014}. Consider a single-cell scenario, when users are clustered into groups according to the similarity of their channel covariance matrices, \cite{adhikary2013} proposed a two-tier precoding approach, where the outer precoder controls inter-group interference based on long-term CSIT (the channel covariance matrices) while the inner precoder controls intra-group interference based on short-term effective channel (the channel cascaded by the outer precoder) with a reduced-dimension. The finding of \cite{adhikary2013} has been generalized into multi-polarized system \cite{Jaehyun2014}. The work \cite{Kim2014} proposed a signal-to-leakage-plus-noise ratio (SLNR)-based outer precoder design and \cite{chen2014} developed a low complexity iterative algorithm to compute the outer precoder. However, the rate performance of the aforementioned two-tier precoding schemes are highly degraded by two limiting factors: inter-group and intra-group interference.

To address these issues in the context of reduced CSI feedback, we propose a novel and general framework, called Hierarchical-Rate-Splitting (HRS), that is particularly suited to FDD massive MIMO systems. By clustering users into groups based on their channel second-order statistics, the proposed HRS scheme exploits the benefits of spatially correlated channels and the two-tier precoding structure. Specifically, HRS is partitioned into an inner RS and an outer RS. Let us imagine each group as a single user, an outer RS tackles the inter-group interference by packing part of one selected user's message into a common codeword that can be decoded by all users. In the presence of multiple users per group, an inner RS copes with the intra-group interference by packing part of one selected user's message into a common codeword that can be decoded by users in that group. Note that the achievable rate of the common message is the minimum rate among users that decode this common message. In contrast to RS where the only one common message should be decoded by all users, HRS offers more benefits by transmitting multiple common messages while the achievable rate of each inner common message is the minimum rate among a smaller number of users.

We analyse the proposed HRS scheme in the large-scale array regime under imperfect CSIT. We also compute a closed-form adaptive power allocation for each message, by which HRS exhibits robustness w.r.t. CSIT error and eigen-subspaces overlap. Moreover, we quantify the sum rate gain of HRS over conventional multiuser broadcasting scheme with two-tier precoder, which offers insights into the effect of system parameters, e.g., SNR, CSIT quality, spatial correlation, etc.

\emph{Organization:} Section \ref{systemmod} introduces the system model. In Section \ref{HRS}, we develop the HRS transmission scheme and elaborate the precoder design, asymptotic rate performance and power allocation. Section \ref{numresults} presents the numerical results and Section \ref{conclusion} concludes the paper. \emph{Notations:} Bold lower and upper case letters denote vectors and matrices, respectively. The notations {\small$[\mathbf{X}]_i,[\mathbf{X}]_{i,j}, \mathbf{X}^T, \mathbf{X}^H, \text{tr}(\mathbf{X}), E(\mathbf{X})$} denote the $i$-th column, the entry in the $i$-th row and $j$-th column, the transpose, conjugate transpose, trace and expectation of a matrix {\small$\mathbf{X}$}. $\|\mathbf{x}\|$ represents the 2-norm of a vector.

\section{SYSTEM MODEL} \label{systemmod}
Consider a single cell FDD downlink system where the BS equipped with $M$ antennas transmits messages to $K (\le M)$ single-antenna users over a spatially correlated Rayleigh-fading channel. Consider a geometrical one-ring scattering model \cite{bruno2013}, the correlation between the channel coefficients of antennas $1 \le i, j \le M$ is given by
{\small
\begin{equation} \label{eq:correlation}
[\mathbf{R}_k]_{i,j} =  \frac{1}{2 \Delta_k} \int^{\theta_k + \Delta_k}_{\theta_k - \Delta_k} e^{-j \frac{2 \pi}{\lambda} \psi(\alpha)(\mathbf{r}_i - \mathbf{r}_j) } d \alpha,
\end{equation}}where $\theta_k$ is the azimuth angle of user $k$ with respect to the orientation perpendicular to the array axis. $\Delta_k$ indicates the angular spread of departure to user $k$. $\psi(\alpha) = [\cos(\alpha), \sin(\alpha) ]$ is the wave vector for a planar wave impinging with the angle of $\alpha$, $\lambda$ is the wavelength and $\mathbf{r}_i = [x_i, y_i]^T$ is the position vector of the $i$-th antenna. With the Karhunen-Loeve model, the downlink channel of user $k$ $\mathbf{h}_k \in \mathbb{C}^{M}$ is expressed as {\small$\mathbf{h}_k =  \mathbf{U}_k \mathbf{\Lambda}^{{1}/{2}}_k \mathbf{g}_k$}, where {\small$\mathbf{\Lambda}_k \in \mathbb{C}^{r_k \times r_k}$} is a diagonal matrix containing the non-zero eigenvalues of the spatial correlation matrix {\small$\mathbf{R}_k$}, and {\small$\mathbf{U}_k \in \mathbb{C}^{M \times r_k}$} consists of the associated eigenvectors. The slowly-varying channel statistics {\small$\mathbf{R}_k$} can be accurately obtained via a rate-limited backhaul link or via uplink-downlink reciprocity and is assumed perfectly known to both BS and users. $\mathbf{g}_k$ has independent and identical distributed (i.i.d.) $\mathcal{CN} (0,1)$ entries.
For each channel use, linear precoding is employed at the BS to support simultaneous downlink transmissions to $K$ users. The received signals can be expressed as
{\small
\begin{equation} \label{eq:rx_ch}
\mathbf{y} = \mathbf{H}^H \mathbf{x} + \mathbf{n},
\end{equation}}where $\mathbf{x} \in \mathbb{C}^{M}$ is the linearly precoded signal vector subject to the transmit power constraint $\mathbb{E}[||\mathbf{x}||^2] \le P$, $\mathbf{H} = [\mathbf{h}_1, \cdots, \mathbf{h}_K]$ is the downlink channel matrix, $\mathbf{n} \sim \mathcal{CN} (\mathbf{0},\mathbf{I}_{K})$ is the additive white Gaussian noise (AWGN) vector and $\mathbf{y} \in \mathbb{C}^{K}$ is the received signal vector at the $K$ users.

\section{HIERARCHICAL RATE SPLITTING} \label{HRS}

\subsection{Transmission Scheme} \label{HRSmodel}
Recently, multiuser broadcasting schemes with a two-tier precoder for FDD massive MIMO systems have been proposed to lessen CSIT requirement by exploiting the knowledge of spatial correlation matrix at the transmitter \cite{adhikary2013, Jaehyun2014, Kim2014, chen2014}. Since the human activity is usually confined in a small region, locations of users tend to be spatially clustered. We make the same assumption as \cite{adhikary2013} that $K$ users are partitioned into $G$ groups (e.g., via K-mean clustering) and that users in each group share the same spatial correlation matrix {\small$\mathbf{R}_g = \mathbf{U}_g \mathbf{\Lambda}_g \mathbf{U}^H_g$} with rank $r_g$. We let $K_g$ denote the number of users in group $g$ such that {\small$\sum^G_{g=1} K_g = K$}. The downlink channel of the $g$-th group is expressed as {\small$\mathbf{H}_g =[\mathbf{h}_{g1}, \cdots, \mathbf{h}_{gK_g}] =  \mathbf{U}_g \mathbf{\Lambda}^{1/2}_g \mathbf{G}_g$}, where the elements of $\mathbf{G}_g$ are distributed with $\mathcal{CN} (0, 1)$. Then, the transmitted signal of conventional two-tier precoded (TTP) broadcasting system is expressed as
{\small
\begin{equation} \label{eq:tx_sigjsdm}
\mathbf{x} = \sum_{g = 1}^{G} \,  \mathbf{B}_{g}  \mathbf{W}_{g} \mathbf{P}_{g}\, \mathbf{s}_{g},
\end{equation}}where {\small$\mathbf{s}_{g} \in \mathbb{C}^{K_g}$} represents the data streams for the $g$-th group users. The outer precoder {\small$\mathbf{B}_g \in \mathbb{C}^{M \times b_g}$} is based on the long-term CSIT while the inner precoder {\small$\mathbf{W}_g \in \mathbb{C}^{b_g \times K_g}$} depends on the short-term effective channel {\small$\bar{\mathbf{H}}_g = \mathbf{B}^H_g \mathbf{H}_g$}. {\small$\mathbf{P}_g \in \mathbb{C}^{K_g \times K_g}$} is the diagonal power allocation matrix with {\small$\mathbf{P}_g = \sqrt{P/K} \cdot \mathbf{I}$}. Then, the received signal of the $k$-th user in g-th group is given by {\small$y_{gk} =  \sqrt{P_{gk}} \mathbf{h}^H_{gk} \mathbf{B}_{g}  \mathbf{w}_{gk} s_{gk} + \sum_{j \ne k}^{K_g}  \sqrt{P_{gj}} \mathbf{h}^H_{gk} \mathbf{B}_{g}  \mathbf{w}_{gj} s_{gj} + \sum_{l \ne g}^{G} \, \mathbf{h}^H_{gk} \mathbf{B}_{l}  \mathbf{W}_{l} \mathbf{P}_{l}\, \mathbf{s}_{l} + n_{gk}$}, where {\small$\mathbf{w}_{gk} = [\mathbf{W}_g]_k$}. To eliminate the inter-group interference, the outer precoder is designed in the nullspace of the eigen-subspace spanned by the dominant eigenvectors of the other groups' spatial correlation matrices. However, the power attached to the weak eigenmodes may leak out to other groups and create inter-group interference. Besides, the intra-group interference cannot be fully be removed due to imperfect CSIT (e.g., limited feedback). To eliminate the interference-limited behavior at high SNR, one can optimize the groups, the users in each group, etc, as a function of the total transmit power and CSIT quality. In general, such an optimization problem is quite complex.

By generalizing the philosophy of RS, we propose a HRS scheme that consists of an outer RS and an inner RS. By treating each group as a single user, an outer RS would tackle the inter-group interference by packing part of one user's message into a common codeword that can be decoded by all users. Likewise, an inner RS would cope with the intra-group interference by packing part of one user's message into a common codeword that can be decoded by multiple users in that group. The common messages are superimposed over the private messages and the transmitted signal of HRS can be written as
{\small
\begin{equation} \label{eq:tx_sig}
\mathbf{x} \!=\! \sqrt{P_{oc}} \mathbf{w}_{oc}  s_{oc} \!+\! \sum_{g = 1}^{G} \mathbf{B}_{g} \!\left(\! \sqrt{P_{ic,g}} \mathbf{w}_{ic,g} s_{ic,g} \!+\! \sqrt{P_{gk}} \mathbf{W}_{g} \mathbf{s}_{g} \!\right),
\end{equation}
}
where $s_{ic,g}$ denotes the inner common message intended to $g$-th group while $s_{oc}$ denotes the outer common message intended to all users. $\mathbf{w}_{ic,g}$ and $\mathbf{w}_{oc}$ are the corresponding unit norm precoding vectors. A uniform power allocation is performed for the private messages and we mainly focus on how to allocate power between the common and private messages. Hence, let $\beta \in (0, 1]$ represent the fraction of the total power that is allocated to the group (inner common and private) messages. Within each group, $\alpha \in (0, 1]$ denotes the fraction of power given to the private messages. Then, the power allocated to each message is jointly determined by $\alpha$ and $\beta$, i.e., $P_{oc} = P (1 - \beta), P_{ic,g} = \frac{P \beta}{G} (1 - \alpha), P_{gk} = \frac{P \beta}{G} \frac{\alpha}{K_g}$. The decoding procedure is performed as follows. Each user sequentially decodes $s_{oc}$ and $s_{ic,g}$, then remove them from the received signal by SIC. The private message intended to each user can be independently decoded by treating all other private messages as noise. By plugging \eqref{eq:tx_sig} into \eqref{eq:rx_ch}, the SINRs of the common messages and the private message of user $k$ are written as $\gamma^{oc}_{gk} = \frac{P_{oc} |\mathbf{h}^H_{gk} \mathbf{w}_{oc} |^2 }{ IN_{gk} }, \gamma^{ic}_{gk} = \frac{P_{ic,g} |\mathbf{h}^H_{gk} \mathbf{B}_g  \mathbf{w}_{ic,g}|^2 }{IN_{gk} - P_{ic,g} |\mathbf{h}^H_{gk} \mathbf{B}_g  \mathbf{w}_{ic,g}|^2}$, $\gamma^p_{gk} = \frac{P_{gk} \, |\mathbf{h}^H_{gk} \mathbf{B}_g \mathbf{w}_{gk}|^2 }{ IN_{gk} - P_{ic,g} \, |\mathbf{h}^H_{gk} \mathbf{B}_g  \mathbf{w}_{ic,g}|^2 - P_{gk} \, |\mathbf{h}^H_{gk} \mathbf{B}_g \mathbf{w}_{gk}|^2 }$, where {\small$IN_{gk}$ $= \sum^G_{l=1} P_{ic,l} \, |\mathbf{h}^H_{gk} \mathbf{B}_l \mathbf{w}_{ic,l} |^2 + \sum^G_{l=1} \sum^{K_g}_{j=1} P_{lj} \, |\mathbf{h}^H_{gk} \mathbf{B}_l \mathbf{w}_{lj}|^2 + 1$}. The achievable rate of the outer common message is given by $R^{HRS}_{oc} = \log_2 (1 + \gamma^{oc})$ with $\gamma^{oc} = \min \{\gamma^{oc}_{gk} ,\; \forall g, k \}$. The sum rate of the inner common messages is given by $R^{HRS}_{ic} = \sum^G_{g = 1} R^{HRS}_{ic,g} =  \sum^G_{g = 1} \log_2 (1 + \gamma^{ic}_g)$ with $\gamma^{ic}_g = \min \{ \gamma^{ic}_{gk}, \forall k \}$. The sum rate of the private messages is given as $R^{HRS}_{p} = \sum^{G}_{g=1} \sum^{K_g}_{k=1} R^{HRS}_{gk} = \sum^{G}_{g=1} \sum^{K_g}_{k=1} \log_2 (1 + \gamma^p_{gk})$. Then, the sum rate of HRS is $R^{HRS}_{\scriptstyle{\text{sum}}} = R^{HRS}_{oc} + R^{HRS}_{ic} + R^{HRS}_{p}$.

\subsection{Precoder Design} \label{HRSprec}
HRS has only access to the channel covariance matrices and the effective channel estimates {\small$\hat{\bar{\mathbf{H}}}_g = \mathbf{B}^H_g \hat{\mathbf{H}}_g$} of dimension $b_g \times K_g$, where {\small $\hat{\mathbf{H}}_g = \mathbf{U}_g \mathbf{\Lambda}^{1/2}_g \, \hat{\mathbf{G}}_g = \mathbf{U}_g \mathbf{\Lambda}^{1/2}_g (\sqrt{1 - \tau^2_g} \, \mathbf{G}_g + \tau_g \mathbf{Z}_g )$} has dimension of $M \times K_g$. Based on long-term CSIT, the outer precoder {\small$\mathbf{B}_{g}$} is designed to eliminate the leakage to other groups. Denoting the number of dominant (most significant) eigenvalues of {\small$\mathbf{R}_g$} by $r^d_g$ and collecting the associated eigenvectors as {\small$\mathbf{U}^d_g \in \mathbb{C}^{M \times r^d_g}$}, we define {\small$\mathbf{U}_{-g} = [\mathbf{U}^d_1, \cdots, \mathbf{U}^d_{g-1}, \mathbf{U}^d_{g+1}, \cdots, \mathbf{U}^d_G ] \in \mathbb{C}^{M \times \sum_{l \ne g} r^d_l }$}. According to the singular value decomposition (SVD), we denote by {\small$\mathbf{E}^{(0)}_{-g}$} the left eigenvectors of {\small$\mathbf{U}_{-g}$} corresponding to the $(M - \sum_{l \ne g} r^d_l)$ vanishing singular values. To reduce the inter-group interference while enhancing the desired signal power, $\mathbf{B}_g$ is designed by concatenating $\mathbf{E}^{(0)}_{-g}$ with the dominant eigenmodes of the covariance matrix of the projected channel {\small$\tilde{\mathbf{H}}_g = (\mathbf{E}^{(0)}_{-g})^H \mathbf{H}_g$}. The covariance matrix is decomposed as {\small$\tilde{\mathbf{R}}_g = (\mathbf{E}^{(0)}_{-g})^H \mathbf{U}_{g} \mathbf{\Lambda}_{g} \mathbf{U}^H_{g} \mathbf{E}^{(0)}_{-g}= \mathbf{F}_{g} \tilde{\mathbf{\Lambda}}_{g} \mathbf{F}^H_{g}$}, where $\mathbf{F}_{g}$ includes the eigenvectors of $\tilde{\mathbf{R}}_g$. Denote {\small$\mathbf{F}^{(1)}_{g}$} as the dominant $b_g$ eigenmodes and then $\mathbf{B}_g$ is given by {\small$\mathbf{B}_g = \mathbf{E}^{(0)}_{-g} \mathbf{F}^{(1)}_{g}$}.

The outer precoder {\small$\mathbf{B}_g$} can be interpreted as being the $b_g$ dominant eigenmodes that are orthogonal to the subspace spanned by the dominant eigen-space of groups $l \ne g$. $b_g$ determines the dimension of the effective channel and should satisfy {\small$K_g \le b_g \le M - \sum_{l \ne g} r^d_l$ and $b_g \le r^d_g$. $r^d_g (\le r_g)$} is a design parameter with a sum rank constraint {\small$\sum^G_{g=1} r^d_g \le M$}.

The inner precoder {\small$\mathbf{W}_g$} can be designed as RZF, i.e., {\small$\mathbf{W}_g = \xi_g \, \hat{\bar{\mathbf{M}}}_g \hat{\bar{\mathbf{H}}}_g$}, where {\small$\hat{\bar{\mathbf{M}}}_g = (\hat{\bar{\mathbf{H}}}_g \hat{\bar{\mathbf{H}}}^H_g + b_g \, \varepsilon \, \mathbf{I}_{b_g} )^{-1}$}. By following \cite{adhikary2013, Jaehyun2014, wagner2012}, the regularization parameter is set as $\varepsilon = {K}/{b P}$ which is equivalent to the MMSE linear filter. $b$ is give by $b = \sum^G_{g=1} b_g$. Then, the power normalization factor is {\small$\xi^2_g = {K_g}/{\text{tr}(\hat{\bar{\mathbf{H}}}^H_g \hat{\bar{\mathbf{M}}}^H_g \mathbf{B}^H_g \mathbf{B}_g \hat{\bar{\mathbf{M}}}_g \hat{\bar{\mathbf{H}}}_g)}$}.

The precoder $\mathbf{w}_{oc} \in \mathbb{C}^{M}$ aims to maximize the achievable rate of the outer common message $\log_2 (1 + \gamma^{oc})$ based on the reduced-dimensional channel estimate {\small$\hat{\bar{\mathbf{H}}}_g \in \mathbb{C}^{b_g \times K_g}, \, \forall g$}. However, there exists a dimension mismatch between $\mathbf{w}_{oc}$ and {\small$\hat{\bar{\mathbf{H}}}_g$}. To address this problem, we first construct {\small$\tilde{\mathbf{H}}_g = \mathbf{B}_g \hat{\bar{\mathbf{H}}}_g \in \mathbb{C}^{M \times K_g}$} and {\small$\tilde{\mathbf{H}} = [\tilde{\mathbf{H}}_1, \cdots, \tilde{\mathbf{H}}_G] \in \mathbb{C}^{M \times K}$}. From \cite[Lemma 5]{wagner2012}, the columns of $\tilde{\mathbf{H}}$ become orthogonal as $M \rightarrow \infty$ and we are able to design the precoder $\mathbf{w}_{oc}$ as a linear combination of $\tilde{\mathbf{h}}_{k} = [\tilde{\mathbf{H}}]_{k}$, which can be interpreted as a weighted matched beamforming (MBF). We simply employ equally weighted MBF in order to obtain a tractable and insightful asymptotic sum rate expression in the sequel.

On the other hand, the optimization of the multiuser transmit precoding is generally a NP-hard problem. Thus, the optimal precoder of the inner common message $\mathbf{w}_{ic,g}$ that maximizes $R^{HRS}_{ic}$ cannot be obtained efficiently. However, when the outer precoder fully eliminates the inter-group interference, $\mathbf{w}_{ic,g}$ can be equivalently designed to maximize $R^{HRS}_{ic,g}$ within each group. Following a similar design philosophy of $\mathbf{w}_{oc}$, we here design $\mathbf{w}_{ic,g}$ as an equally weighted MBF of the effective channel {\small$\hat{\bar{\mathbf{H}}}_g$}. Under further assumption that $K \rightarrow \infty$, we note that {\small$\hat{\bar{\mathbf{M}}}_g$} of the inner precoder {\small$\mathbf{W}_g( = \xi_g \, \hat{\bar{\mathbf{M}}}_g \hat{\bar{\mathbf{H}}}_g)$} can be approximated by an identity matrix. Hence, $\mathbf{w}_{ic,g}$ can be equivalently designed as an equally weighted MBF of {\small$\mathbf{W}_g$}, i.e.,$\mathbf{w}_{ic,g} = \zeta_{ic,g} \hat{\bar{\mathbf{q}}}_g$, where $\hat{\bar{\mathbf{q}}}_g =  \frac{1}{K_g}\sum^{K_g}_{k = 1} \mathbf{w}_{gk}$ and $\zeta^2_{ic,g} = {1}/{\hat{\bar{\mathbf{q}}}^H_g \mathbf{B}^H_g \mathbf{B}_g \hat{\bar{\mathbf{q}}}_g}$.

\subsection{Asymptotic Rate Analysis} \label{HRSperf}
For simplicity of exposition, we assume that $\tau_g = \tau, K_g = \bar{K}, b_g = \bar{b}, \forall g$. We shall omit the proof of the following theorem, where the asymptotic SINRs of HRS can be directly established based on the approach of \cite{wagner2012}.

$\textbf{Theorem 1:}$ As $M, K, b \rightarrow \infty$ with fixed ratios $\frac{K}{M}$ and $\frac{b}{M}$, the SINRs of HRS asymptotically converge as
\begin{equation} \label{eq:desnr1}
\begin{array}{lcl}
\gamma^{oc}_{gk}  - \gamma^{oc,\circ}_{g} \rightarrow  0,  \;
\gamma^{ic}_{gk}  - \gamma^{ic,\circ}_{g} \rightarrow  0, \;
\gamma^{p}_{gk} - \gamma^{p,\circ}_{g} \rightarrow  0,
\end{array}
\end{equation}
almost surely, where
{\small
\begin{equation} \label{eq:desnr2}
\gamma^{oc,\circ}_{g} = \frac{ \kappa_g P(1 - \beta) (1 - \tau_g^2)} {\beta \big(\sum_{l \neq g} (\xi^{\circ}_{l})^2 \Upsilon^{\circ}_{gl} + \big(\xi^{\circ}_{g} \big)^2 \Upsilon^{\circ}_{gg} \Omega_g + \frac{P}{K} \big(\xi^{\circ}_{g} \big)^2 \Phi_g \big) + 1}
\end{equation}
\begin{equation} \label{eq:desnr3}
\gamma^{ic,\circ}_{g} = \frac{\beta (1 - \alpha) \big(\xi^{\circ}_{g} \big)^2 \left(\Upsilon^{\circ}_{gg} \Omega_g + \frac{P}{K} \Phi_g \right) }{\beta \sum_{l \neq g} \left(\xi^{\circ}_{l}\right)^2 \Upsilon^{\circ}_{gl} + \beta \alpha \, \big(\xi^{\circ}_{g} \big)^2 (\Upsilon^{\circ}_{gg} \Omega_g + \frac{P}{K} \Phi_g ) + 1},
\end{equation}
\begin{equation} \label{eq:desnr4}
\gamma^{p,\circ}_{g} = \frac{\beta \alpha \frac{P}{K} \big(\xi^{\circ}_{g} \big)^2 \Phi_g}{\beta \sum_{l \neq g} \left(\xi^{\circ}_{l}\right)^2 \Upsilon^{\circ}_{gl} + \beta \alpha \big(\xi^{\circ}_{g} \big)^2 \Upsilon^{\circ}_{gg} \Omega_g + 1},
\end{equation}}with {\small$(\xi^{\circ}_{g})^2 = \frac{\bar{K}}{\Psi^{\circ}_g}, \, \Psi^{\circ}_g = \frac{\bar{K}}{\bar{b}} \frac{ m'_{g}} { (1 + m^{\circ}_{g})^2}, \, \Phi_g = \frac{ (1 - \tau^2_g) (m^{\circ}_g)^2}{(1 + m^{\circ}_g)^2}, \, \Upsilon^{\circ}_{gl} = \frac{P}{G} \frac{1}{\bar{b}} \frac{m'_{gl}}{(1 + m^{\circ}_{l})^2}, \, \kappa_g = \frac{\text{Tr}(\bar{\mathbf{R}}_{gg})^2}{\bar{K} \sum^G_{l = 1} \text{Tr}(\bar{\mathbf{R}}_{ll})}, \, \bar{\mathbf{R}}_{gl} = \mathbf{B}^H_l \mathbf{R}_g \mathbf{B}_l, \; \forall g, l, \, \Omega_g = \frac{\bar{K}-1}{\bar{K}}  \frac{(1 - \tau^2_g (1 - (1 + m^{\circ}_g)^2)) }{(1 + m^{\circ}_g)^2}$}, and
{\small
\begin{equation} \label{eq:desub31}
m'_{g} = \frac{\frac{1}{\bar{b}} \text{Tr}(\bar{\mathbf{R}}_{gg} \mathbf{T}_g \mathbf{B}^H_g \mathbf{B}_g \mathbf{T}_g) }{1 - \frac{\frac{\bar{K}}{\bar{b}} \text{Tr}(\bar{\mathbf{R}}_{gg} \mathbf{T}_g \bar{\mathbf{R}}_{gg} \mathbf{T}_g) }{\bar{b} \, (1 + m^{\circ}_{g})^2 } },
m'_{gl} = \frac{\frac{1}{\bar{b}} \text{Tr}(\bar{\mathbf{R}}_{ll} \mathbf{T}_l \bar{\mathbf{R}}_{gl} \mathbf{T}_l) }{1 - \frac{\frac{\bar{K}}{\bar{b}} \text{Tr}(\bar{\mathbf{R}}_{ll} \mathbf{T}_l \bar{\mathbf{R}}_{ll} \mathbf{T}_l) }{\bar{b} \, (1 + m^{\circ}_{l})^2 } },\nonumber
\end{equation}}where $m^{\circ}_{g}$ and $\mathbf{T}_g$ the unique solutions of the following: {\small $m^{\circ}_{g} = \frac{1}{\bar{b}} \text{Tr}(\bar{\mathbf{R}}_{gg} \mathbf{T}_g ), \, \mathbf{T}_g = ( \frac{\bar{K}}{\bar{b}} \frac{\bar{\mathbf{R}}_{gg}}{1 + m^{\circ}_{g}} + \varepsilon \, \mathbf{I}_{\bar{b}} )^{-1}$}.

It follows from \eqref{eq:desnr1} that {\small$(R^{HRS}_p \!-\! R^{HRS,\circ}_p)/K \!\overset{M \rightarrow \infty}{\longrightarrow}\! 0$} where {\small$R^{HRS,\circ}_p = \sum^G_{g=1} \bar{K} \log_2 (1 + \gamma^{p, \circ}_{g})$, $(R^{HRS}_{ic} \!-\! R^{HRS,\circ}_{ic})/G \overset{M \rightarrow \infty}{\longrightarrow} 0$} where {\small$R^{HRS,\circ}_{ic} = \sum^G_{g=1} \log_2 (1 + \gamma^{ic, \circ}_{g})$}, and that {\small$R^{HRS}_{oc} - R^{HRS,\circ}_{oc} \overset{M \rightarrow \infty}{\longrightarrow} 0$} where {\small$R^{HRS,\circ}_{oc} = \log_2 (1 + \gamma^{oc, \circ})$} with {\small$\gamma^{oc,\circ} = \text{min} \{\gamma^{oc,\circ}_g, \forall g\}$}. Then, an approximation {\small$R^{HRS,\circ}_{\scriptstyle{\text{sum}}}$} of the sum rate of HRS is obtained as {\small$R^{HRS,\circ}_{\scriptstyle{\text{sum}}} = R^{HRS,\circ}_{oc} + R^{HRS,\circ}_{ic} + R^{HRS,\circ}_p$}.

Likewise, the asymptotic sum rate of the conventional TTP in \eqref{eq:tx_sigjsdm} converges as {\small$(R^{TTP}_{\scriptstyle{\text{sum}}}- R^{TTP, \circ}_{\scriptstyle{\text{sum}}})/K \!\overset{M \rightarrow \infty}{\longrightarrow}\! 0$}, where {\small$R^{TTP, \circ}_{\scriptstyle{\text{sum}}} = \sum^G_{g=1} \bar{K} \log_2 (1 + \gamma^{TTP, \circ}_{g})$} and
{\small
\begin{equation} \label{eq:snrjsdm}
\gamma^{TTP, \circ}_{g} = \frac{\frac{P}{K} \big(\xi^{\circ}_{g} \big)^2 \Phi_g}{\sum_{l \neq g} \left(\xi^{\circ}_{l}\right)^2 \Upsilon^{\circ}_{gl} + \big(\xi^{\circ}_{g} \big)^2 \Upsilon^{\circ}_{gg} \Omega_g + 1},
\end{equation}
}and the first term in the denominator of \eqref{eq:snrjsdm} containing {\small$\Upsilon^{\circ}_{gl} (\bar{\mathbf{R}}_{gl})$} denotes inter-group interference while the second term with {\small$\Omega_g (\tau^2)$} refers to intra-group interference. The sum rate gain of HRS over conventional two-tier precoding BC is quantified by {\small$\Delta R^{HRS,\circ} = R^{HRS,\circ}_{oc} + R^{HRS,\circ}_{ic} + \sum^G_{g = 1} \bar{K} \big(\log_2 \big(1 + \gamma^{p,\circ}_{g}\big) - \log_2 \big(1 + \gamma^{TTP,\circ}_{g} \big) \big)$}.

\subsection{Power Allocation} \label{HRSpower}
Since $\alpha$ and $\beta$ are coupled in the SINR expressions \eqref{eq:desnr2} $\sim$ \eqref{eq:desnr4},  a closed-form and optimal solution that maximizes the sum rate of HRS {\small$R^{HRS,\circ}_{\scriptstyle{\text{sum}}}$} cannot be obtained in general. In this paper, we compute a closed-form suboptimal but effective power allocation method, by which the private messages of HRS are allocated a fraction of the total power and achieve nearly the same sum rate as the conventional broadcasting scheme with full power, i.e., {\small $R^{TTP,\circ}_{\scriptstyle{\text{sum}}} \approx R^{HRS,\circ}_p$}. Then, the remaining power is utilized to transmit the common messages and therefore enhance performance. We can write
{\small
\begin{equation} \label{eq:poweralc1}
\gamma^{p,\circ}_g \le \gamma^{TTP, \circ}_g, \quad \forall g,
\end{equation}
}for $\forall \alpha, \beta \in (0, 1]$. Consider two extreme cases: weak and strong inter-group interference. Based on \eqref{eq:snrjsdm}, the notation of `weak' implies that the inter-group interference is sufficiently small and therefore can be negligible, i.e., {\small$\Upsilon^{\circ}_{gl} \approx 0, \forall g \ne l$}. The sum rate {\small$R^{TTP,\circ}_{\scriptstyle{\text{sum}}}$} is limited by the intra-group interference due to imperfect CSIT. On the contrary, the notation of `strong' means that the inter-group interference dominates the rate performance, i.e., {\small$\sum_{l \neq g} \left(\xi^{\circ}_{l}\right)^2 \Upsilon^{\circ}_{gl} > \big(\xi^{\circ}_{g} \big)^2 \Upsilon^{\circ}_{gg}$}.

$\textbf{Proposition 1:}$ The equality of \eqref{eq:poweralc1} nearly holds when the power splitting ratios $\alpha$, $\beta$ are given as
{\small
\begin{equation} \label{eq:poweralc2}
\beta = 1, \quad \alpha = \text{min} \; \Big\{ \frac{\bar{K}}{P \cdot \Gamma_{IG}}, \; 1 \Big\}
\end{equation}}in the weak inter-group interference regime, and as
{\small
\begin{equation} \label{eq:poweralc22}
\beta = \text{min} \; \Big\{\frac{K}{P \cdot \Gamma_{OG} + \bar{K}}, \; 1 \Big\}, \quad \alpha = 1
\end{equation}}in the strong inter-group interference regime, where
{\small
\begin{eqnarray} \label{eq:poweralc3}
\Gamma_{OG} &=& \underset{g}{\text{min}} \; \bigg\{ \sum_{l \ne g} \frac{1}{G} \frac{\text{tr}\big(\bar{\mathbf{R}}_{gl} \bar{\mathbf{R}}^{-1}_{ll}\big)}{\text{tr}\big(\bar{\mathbf{R}}^{-1}_{ll}\big)} \bigg\}, \\
\Gamma_{IG} &=& \underset{g}{\text{min}} \; \bigg\{ \frac{\tau^2}{G} \frac{\bar{b}}{\text{tr}\big(\bar{\mathbf{R}}^{-1}_{gg}\big)} \frac{\bar{K}-1}{\bar{K}} \bigg\}.
\end{eqnarray}
}
\begin{IEEEproof}
See Appendix B in \cite{dai2015}.
\end{IEEEproof}

When the inter-group interference is negligible, HRS becomes a set of parallel RS in each group, i.e., the outer common message is unnecessary. By contrast, when the inter-group interference is the dominant degrading factor, the inner common message transmission as well as the private messages transmission are inter-group interference limited. In this case, HRS boils down to RS (with reduced-dimensional CSIT). For the general inter-group interference case, finding a closed-form solution of $\alpha, \beta$ that guarantees a sum rate gain of HRS over two-tier precoding BC is challenging. Nevertheless, motivated by the extreme cases in Proposition 1, we propose the following power allocation strategy. $\alpha$ is determined by \eqref{eq:poweralc2} while $\beta$ comes from \eqref{eq:poweralc22}. Then, $\alpha$ is reset to 1 if $\beta < 1$ in order to guarantee the equality in \eqref{eq:poweralc1}. Simulation results show that the proposed power allocation strategy works well in any inter-group interference regime. Thus, the BS can readily compute the power allocation based on long-term CSIT.

From \eqref{eq:poweralc2} and \eqref{eq:poweralc22}, we have {\small$\alpha = \beta = 1$} at low SNR and HRS becomes the conventional two-tier precoding BC, leading to {\small$\Delta R^{HRS,\circ} = 0$}. Namely, the effect of imperfect CSIT/overlapping eigen-subspaces on the sum rate of broadcasting private messages is negligible and thereby common message(s) is not needed. On the other hand, the rate performance of the conventional two-tier precoding BC schemes saturates at high SNR while HRS exploits a fraction of the total power ($\alpha < 1$ or $\beta < 1$) to transmit the common message(s) and enhance the sum rate.

$\textbf{Corollary 1:}$ With power allocation of Proposition 3, the sum rate gain $\Delta R^{HRS,\circ}$ at high SNR is lower bounded as:
{\small
\begin{equation} \label{eq:hrsgain1}
\Delta R^{HRS,\circ}  \ge \sum^G_{g=1} \big(\log_2 (1 + \gamma^{ic, \circ}_{g}) - \log_2 (e)\big),
\end{equation}}in the weak inter-group interference regime, and as
{\small
\begin{equation} \label{eq:hrsgain2}
\Delta R^{HRS,\circ}  \ge \log_2 (1 + \gamma^{oc, \circ}) - \log_2 (e),
\end{equation}}in the strong inter-group interference regime.
\begin{IEEEproof}
By plugging \eqref{eq:poweralc2} and \eqref{eq:poweralc22} into {\small$\Delta R^{HRS,\circ}$}, we upper bound {\small$ R^{TTP,\circ}_{\scriptstyle{\text{sum}}} - R^{HRS,\circ}_p$} at high SNR and obtain \eqref{eq:hrsgain1} and \eqref{eq:hrsgain2}. More details are given in Appendix C of \cite{dai2015}.
\end{IEEEproof}

\emph{Remark 1:} The following are some physical interpretations of Proposition 1 and Corollary 1.

\textbf{Power allocation to the private and common messages}: The intra-group power splitting ratio $(\alpha)$ decreases as $\tau^2$ increases. Namely, in order to alleviate intra-group interference, we should allocate less power to the private messages as the CSIT quality gets worse $(\tau^2 \rightarrow 1)$. Similarly, the inter-group power splitting ratio $(\beta)$ drops as the inter-group interference term $\Upsilon^{\circ}_{gl}, g \ne l$ becomes larger. From \eqref{eq:poweralc2} $\sim$ \eqref{eq:poweralc22}, the power distributed to the privates messages is an invariant of $P$ at high SNR: {\small$\sum^G_{g=1}\sum^{\bar{K}}_{k=1} P_{gk} = P \alpha \beta = \left\{
\begin{array}{lcl}
\frac{K}{\Gamma_{OG}} , \quad \text{if} \; \beta < 1 \\
\frac{\bar{K}}{\Gamma_{IG}} , \quad \text{otherwise}
\end{array}
\right.$}, which places the sum rate of private messages back to the non-interference-limited regime. Meanwhile, the power allocated to the common messages linearly increases with $P$ at high SNR.

\textbf{Sum rate gain}: HRS exploits the extra power beyond saturation of conventional broadcasting schemes to transmit the common messages, leading to a sum rate that increases with the available transmit power. In the weak inter-group interference regime, HRS becomes a set of parallel inner RS. Based on \eqref{eq:hrsgain1}, the sum rate gain {\small$\Delta R^{HRS,\circ}$} increases by $G$ bps/Hz for each 3 dB power increment at high SNR. By contrast, HRS boils down to RS in the strong inter-group interference regime and {\small$\Delta R^{HRS,\circ}$} increases by $1$ bps/Hz for each 3 dB power increment at high SNR.

\section{SIMULATION RESULTS} \label{numresults}
Numerical results are provided to validate the effectiveness of RS and HRS. Uniform circular array (UCA) with $M = 100$ isotropic antennas are equipped at the BS. Consider the transmit correlation model in \eqref{eq:correlation}, the antenna elements are equally spaced on a circle of radius $\lambda D$, for {\small$D = {0.5}/{\sqrt{(1-\cos(2 \pi/M))^2 + \sin(2 \pi/M)^2}}$}, leading to a minimum distance $\lambda/2$ between any two antennas.

Consider $K = 12$ users equally clustered into $G = 4$ groups. We compare the proposed HRS scheme with the following baselines: \textbf{Baseline 1 (BC with two-tier precoder \cite{adhikary2013})}, \textbf{Baseline 2 (Baseline 1 with user scheduling at the group level)}: Within each group, a single user with the largest effective channel gain is selected and the precoder of the private message intended to each user is MBF. \textbf{Baseline 3 (Baseline 1 with user scheduling at the system level)}: User scheduling is performed at the system level such that the best user among all is selected. Two types of HRS are investigated: exhaustive search (HRS\_EXS) and closed-form (HRS\_CLF). Specifically, HRS\_EXS performs a simulation-based exhaustive search with step 0.01 for the best power splitting ratios $\alpha$ and $\beta$. HRS\_CLF allocates power by following the closed-form solution in Proposition 1.

Various CSIT qualities have been simulated and $\tau^2 = 0.4$ is taken as examples. For the outer precoder design, we set $\bar{b} = 15$ such that $\bar{K} \le \bar{b} \le M - (G-1) r^d$ and $\bar{b} \le r^d$, where $r^d = 20$ includes the dominant eigenvalues of $\mathbf{R}_g, \forall \, g$. To verify the effectiveness of the proposed HRS strategy, we consider two scenarios with disjoint and overlapping eigen-subspaces, respectively. As an example, we set $\theta_g = -\frac{\pi}{2} + \frac{\pi}{3}(g-1)$ and $\Delta_g = \Delta = \frac{\pi}{8}, \forall \,g$ corresponding to disjoint eigen-subspaces $([\theta_g - \Delta_g, \theta_g + \Delta_g] \bigcap [\theta_l - \Delta_l, \theta_l + \Delta_l] = \emptyset, \forall \, l \ne g)$ while $\Delta_g = \Delta = \frac{\pi}{3}, \forall \,g$ leading to eigen-subspaces overlap.

\begin{figure}[t]
  \centering
\subfigure[]{\label{fig_gain1} \includegraphics[width = 0.34\textwidth, height = 0.25\textwidth]{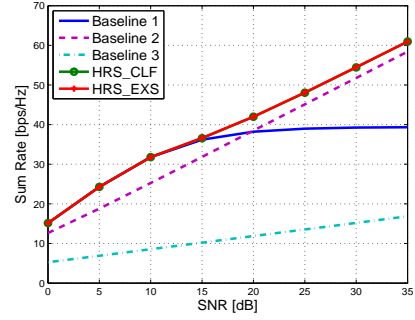}} \\ \vspace{-10pt}
\subfigure[]{\label{fig_gain2} \includegraphics[width = 0.34\textwidth, height = 0.25\textwidth]{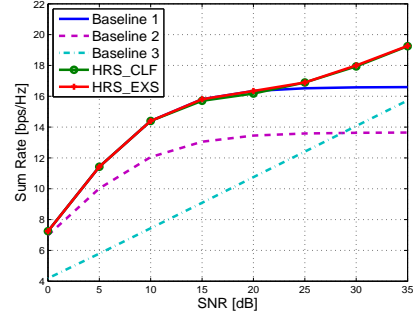}}
\caption{HRS vs. various baselines under imperfect CSIT. (a) disjoint eigen-subspace. (b) overlapping eigen-subspace.} \label{fig_gain12}
\end{figure}

The benefits of HRS under imperfect CSIT are evaluated. With disjoint eigen-subspaces (negligible inter-group interference), Fig. \ref{fig_gain1} shows that conventional multiuser broadcasting scheme with two-tier precoder (Baseline 1) saturates at high SNR due to intra-group interference while user scheduling enables a multiplexing gain of 4 (Baseline 2) and 1 (Baseline 3), respectively. According to Proposition 1, HRS becomes a set of parallel inner RS. We observe that the proposed HRS scheme exhibits substantial rate gain over various baselines. For instance, the sum rate gain of HRS {\small$\Delta R^{HRS}$} over two-tier precoding BC at SNR = 30 dB is 15.5 bps/Hz. With severely overlapping eigen-subspaces (strong inter-group interference), HRS boils down to RS (with reduced-dimensional CSIT) at the system level according to Proposition 1, i.e., inner common messages are not transmitted. Fig. \ref{fig_gain2} reveals that HRS outperforms two-tier precoding BC with/without user scheduling. The sum rate enhancement of HRS over two-tier precoding BC at SNR = 30 dB is 1.5 bps/Hz.

Interestingly, in both settings of Fig. \ref{fig_gain1} and Fig. \ref{fig_gain2}, the closed-form power allocation achieves almost the same sum rate as that of a simulation-based exhaustive search. This verifies the effectiveness of the power allocation strategy in Proposition 1. In Fig. \ref{fig_gain1} and Fig. \ref{fig_gain2}, respectively, we observe that the sum rate gain {\small$\Delta R^{HRS}$} of HRS over two-tier precoding BC increases by nearly $G$ and 1 bps/Hz for any 3 dB increment of power at high SNR. This observation verifies the discussion of Remark 1. In a nutshell, HRS exhibits robustness w.r.t. CSIT error and eigen-subspaces overlap.

\section{CONCLUSION} \label{conclusion}

Due to imperfect CSIT, the rate performance of conventional multiuser broadcasting schemes is severely degraded. To tackle the multiuser interference, we proposed a novel Hierarchical Rate Splitting strategy which exploits the channel second-order statistics and a two-tier precoding structure. Particularly, on top of the private messages, HRS transmits an outer common message and multiple inner common messages that can be decoded by all users and a subset of users, respectively. The outer common message tackles the inter-group interference due to overlapping eigen-subspaces while the inner common messages helps with mitigating the intra-group interference due to imperfect CSIT. Simulation results showed that the proposed HRS strategy achieves significant sum rate gain over the conventional broadcasting schemes with two-tier precoder and HRS exhibit robustness w.r.t. CSIT error and eigen-subspaces overlaps.


\bibliographystyle{IEEEtran}
\bibliography{reference}

\end{document}